\documentclass{article}

\usepackage[letterpaper,top=2cm,bottom=2cm,left=3cm,right=3cm,marginparwidth=1.75cm]{geometry}

\usepackage{amsmath}
\usepackage{graphicx}
\usepackage[colorlinks=true, allcolors=blue]{hyperref}

\begin{document}

\noindent
{\bf{\Large{Roman Early-Definition Astrophysics Survey Opportunity:}}}
\begin{center} 
{\bf{\Large{Galactic Roman Infrared Plane Survey (GRIPS)}}}\\
\vspace*{0.5truecm}
{\Large{White Paper Submitted on October 22 2021}}
\end{center}

\vspace*{1truecm}
\noindent
{{\bf{\Large{Authors:}}} Roberta Paladini (Caltech-IPAC), Catherine Zucker (STScI), Robert Benjamin (Wisconsin), David Nataf (JHU), Dante Minniti (Univ Andres Bello), Gail Zasowski (Univ of Utah), Joshua Peek (STScI), Sean Carey (Caltech-IPAC), Lori Allen (NOIRLab), Javier Alonso-Garcia (Univ Antofagasta), Joao Alves (Univ of Vienna), Friederich Anders (UNiv of Barcelona), Evangelie Athanassoula (LAM), Timothy C. Beers (Univ of Notre Dame),  Jonathan Bird (Vanderbilt Univ),  Joss Bland-Hwathorn (Univ of Sydney), Anthony Brown (Univ of Leiden), Sven Buder (ANU), Luca Casagrande (ANU), Andrew Casey (Monash Univ), Santi Cassisi (INAF), Marcio Catelan (PUC), Ranga-Ram Chary (Caltech-IPAC), Andre-Nicolas Chene (Gemini Obs), David Ciardi (Caltech-IPAC), Fernando Comeron (ESO), Roger Cohen (STScI), Thomas Dame (SAO), Ronald Drimmel (INAF), Jose Fernandez Trincado (UCN), Douglas Finkbeiner (Harvard Univ), Douglas Geisler (Univ de Concepcion), Mario Gennaro (STScI), Alyssa Goodman (Harvard Univ), Gregory Green (MPIA), Gergely Hajdu (CAMK), Calen Henderson (Caltech-IPAC), Joseph Hora (CfA), Valentin D. Ivanov (ESO), Davy Kirkpatrick (Caltech-IPAC), Chiaki Kobayashi (UNiv of Hertfordshire), Michael Kuhn (Univ of Hertfordshire), Andres Kunder (Saint Martin's Univ), Jessica Lu (UC Berkeley), Philip W. Lucas (Univ of Hertfordshire), Daniel Majaess (MSVU), S. Thomas Megeath (Univ of Toledo), Aaron Meisner (NOIRLab), Sergio Molinari (INAF), Przemek Mroz (Warsaw Univ), Meliss Ness (Columbia Univ), Nadine Neumayer (MPIA), Francisco Nogueras-Lara (MPIA), Alberto Noriega-Crespo (STScI), Radek Poleski (Warsaw Univ), Hans-Walter Rix (MPIA), Luisa Rebull (Caltech-IPAC), Henrique Reggiani (Carnegie Obs), Marina Rejkuba (ESO), Roberto K. Saito (UFSC), Ralph Schoenrich (UNiv College London), Andrew Saydjari (Harvard Univ), Eugenio Schisano (INAF), Edward Schlafly (STScI), Keving Schlaufman (JHU), Leigh Smith (Cambridge Univ), Joshua Speagle (Univ Toronto), Dan Wisz (UC Berkeley), Rosemary Wyse (JHU), Nadia Zakamska (JHU)

\vspace*{0.5truecm}
\noindent
{\bf{\Large{Do you support the selection of a Roman Early-Definition Astrophysics Survey?}}}\\

\noindent
Yes. A wide-field near-infrared survey of the Galactic disk and bulge/bar(s) is supported by a large representation of the community of Galactic astronomers. The combination of sensitivity, angular resolution and large field of view make Roman uniquely able to study the crowded and highly extincted lines of sight in the Galactic plane. A $\sim$ 1000 deg$^{2}$ survey of the bulge and inner Galactic disk would yield an impressive dataset of $\sim$120 billion sources and map the structure of our Galaxy. The effort would foster subsequent expansions in numerous dimensions (spatial, depth, wavelengths, epochs).Importantly, the survey would benefit from early defintion by the community, namely because the Galactic disk is a complex environment, and different science goals will require trade offs. \\

\noindent
{\bf{\Large{Science Investigation:}}} The Milky Way is the only large galaxy where individual stars can be resolved down to the central few parsecs. Existing large-scale photometric, spectroscopic, and astrometric surveys have fostered a rich understanding of the Galaxy as a spatially, chemically, and kinematically complex structure, with ample evidence of interactions, past mergers, and secular evolution and substructure in star formation. Recent abundance of large-scale ground-based spectroscopic surveys, and measurements of parallaxes and proper motions by the Gaia mission have super-charged these investigations. A wide area Galactic survey with Roman will characterize most of the stellar content of our Galaxy and will provide unique information on both the history of galaxy formation, and the on-going process of star formation in vastly different environments, as Roman is uniquely suited to deal with the confusion and extinction prevalent in the plan of the Galaxy (see Fig.~1). A Galactic Plane survey was one of five programs specifically endorsed by the Science Definition Team (SDT) in the WFIRST Interim Report (Green et al. 2012). Importanty, the Nancy Grace Roman Space Telescope significantly exceeds the capability of the existing efforts in three critical areas: {\bf{(a) astrometric precision, (b) survey sensitivity, (c) maping speed.}} This opens new avenues in studies of stellar astrophysics, star formation and Galactic structure, e.g. Appendix D of the WFIRST-AFTA report (2015) and Stauffer et al. (2018).\\

\noindent
The {\bf{high angular resolution}} of Roman will enable studies of previously unresolved stellar populations (see Fig.~2). That includes globular clusters in the Galactic plane and bulge, stellar clusters in star forming regions, and the entire nuclear region of the Galaxy. The Nobel-prize winning study of stellar motions near Sgr A$\*$, the HST Galactic center study (200 mas, 0.1 deg$^{2}$, Dong et al. 2011) and the ESO VLT GALACTICNUCLEUS survey (220 mas, 0.3 deg$^{2}$, Nogueras-Lara et al. 2019) are all pertinent examples highlightining the relevance of such a dataset. The {\bf{sensitivity}} of Roman will provide the deepest infrared Galactic plane survey by at least two magnitudes (see Fig.~3 and Fig.~4). Red clumps and YSOs can be surveyed out to a greater volume of the disk allowing the rewriting of Galactic structure, particularly the spiral arms and the central Galaxy where source confusion has blocked progress. The greater depth will likewise enable studies of the stellar initial mass function down to lower mass limits in sites across the Galaxy, and provide significantly more ``background" sources for the construction of 3D dust maps. The combination of depth and angular resolution wil also yiel a novel/unique catalog of galaxies and galaxy clusters beyond the Galactic disk. The {\bf{mapping speed}} of Roman will allow for a significant graction of the stars in the Galaxy to be covered in a uniform way, a crucial requiremnt for studies of Galactic structure. Finally, a Roman single-pass survey of the Galactic Plane early in the mission would enable subsequent passes later on, largely surpassing what can be obtained by simply combining Roman and, e.g., 2MASS, with a 25-year baseline, therefore bolstering the characterization of stellar proper motions in regions inaccessible to Gaia, notably in the complex orbital structure of the Galactic bar(s) and nucleus. This will produce new insights on the ``inside-out" evolution and central luminous/dark matter distribution of the Galaxy, and enable proper motion selection of populations, e.g., HST SWEEPS survey. GRIPS (Galactic Roman Infrared Plane Survey) will also allow for synergies with shorter wavelength monitoring of the Galactic Plane by the Vera C. Rubin Observatory, with the spectral information obtained by SPHEREx and SDSS-V Milky Way Mapper, and with the proposed (Hobbs et al. 2016) Gaia-NIT mission.\\ 

\noindent
{\bf{Possible Observational Outline and Preparatory Activities:}} To maximize the impact of Roman's high angular resolution in the Galaxy's most crowded fields, we propose a 991 deg$^{2}$ survey of the inner Galactic Plane, spanning latitudes $|b| <$ 3$^{2}$ over the longitude range $|l| <$ 60$^{2}$ with additional latitude coverage up to $|b| <$ 10$^{2}$ in the bulge ($|l| <$ 10 $^{2}$). We will leverage the Wide Field Instrument in three filters: F106, F158 and F213. The F106 filter was chosen to provide continuous wavelength coverage with Rubin at shorter wavelengths, and the F213 filter was selected to maximize the potential of Roman in dust-enshrouded regions deep in the plane. F158 will complement the other two filters and allow building diagnostics for the identification of the surveyed stellar populations. Importantly, follow-up time in subsequent years will allow additional astrometric and proper motion measurements over a sizable temporal baeline.\\

\noindent
We propose an integration time of 55 seconds per filter, reaching a minimum depth of 25.5 mag in F106, 25.3 mag in F158, and 24.7 mag in F213. We plan for one primary dither in each filter to fill the gaps in the detectors and account for cosmic rays - totally 21.4 sec- and two secondary sub-pixel dithers only in the F213 band to obtain accurate astrometry for determining the proper motions, requiring 10 sec each for slewing and settling. This yields two exposures each at F106 and F158, and six exposures in F213. We propose small FOV-type slews (of 0.4$^{\circ}$), which will add 50 sec of overhead for each field. With a 3 sec readout time between exposures when not slewing, this setup will require 673 sec of time per 0.281 deg$^{2}$ field, and includes exposure time, slewing time, readout time, and time for our primary (gap-fill) and secondary (sub-pixel) dithering strategies. We will need approximately 3600 pointings for our 991 deg$^{2}$ survey area, yielding and estimated total time of 673 hours. By extrapolating to our proposed footprint the Penny et al. (2019) stellar density estimates based on the Besancon model, we estimate that up to 120 billion unique stellar sources shall be characterized, as compared to the 0.38 billion sources in Gaia eDR3.\\

\noindent
{\em{Optimization of Survey Design via Community Inputs:}} We anticipate that the ultimate design of GRIPS wil require substantial community input to optimize the undertaking, as highlighted by the breadth of expertise in the list of co-authors. The above survey design is intended to serve as a baseline, with trade-offs in the dithering strategy, coverage area, and number/choice of filters having substantial implications for the different science cases described in Section 4. Defining the survey early will allow us to build not only the most powerful survey to address these different science cases on its own, but also a well-crafted initial design to enable optimal expansions in epochs, spatial coverage, wavelength coverage by subsequent guest investigators. A poorly-considered initial design will foreclose some of these options.\\

\noindent{\em{Development of A Crowded-Fields Forced Photometry Pipeline:}} Creating an expandable high precision astro-/photometric catalog -- with $\sim$ 100 times more sources than Gaia -- in regions of spatially complex diffuse background will require significant preparatory work. Combining with next generation ground-based surveys like Rubin and existing longer wavelength surveys will allow us to significantly expand the native Roman photometric coverage, broadening the scientific reach of the survey. The development of PSF forced photometry pipelines able to operate in heavily crowded fields is required given the resolution of ground-based observations. This activity, which we will carry out during the early definition period using the combined HST/DECam dataset towards the Galactic center, will complement effort such as the Joint Survey Processing (Chary et al. 2020), which is focused on lower source-density regions.\\

\bibliographystyle{alpha}
\bibliography{sample}

\noindent
Chary et al., 2020, {\em{Joint Survey Processing of Euclid, Roman and Rubin: Final Report}}, astro-ph/2008.10663\\

\noindent
Dong et al., 2011, MNRAS, 417, 114\\

\noindent
Green et al. 2012, {\em{Wide-Field Infrared Survey Telescope (WFIRST): Final Report}}, astro-ph/1208.4012\\

\noindent
Hobbs et al., 2016, {\em{Gaia NIR: Combining optical and Near-Infrared (NIR) capabilities with Time-Delay Integration (TDI) sensors for a future Gaia-like mission}}, astro-ph/1609.07325\\

\noindent
Minniti et al., 2010, NewAstr, 15, 433\\

\noindent
Nogueras-Lara et al., 2019, A$\&$A, 631, 20\\

\noindent
Penny et al., 2019, ApJS, 241, 3\\

\noindent
Spergel et al., 2015, {\em{Wide-Field Infrared Survey Telescope-Astrophysics Focused Telescope Assets}}, WFIRST-AFTA 2015 Report\\

\noindent
Stauffer et al., 2018, {\em{The Science Advantage of a Redder Filter for WFIRST}}, astro-ph/1806.00554\\

\vspace*{1truecm}
\noindent
{\Large{\bf{Figures}}}

\begin{figure}
\centering
\includegraphics[width=1\linewidth]{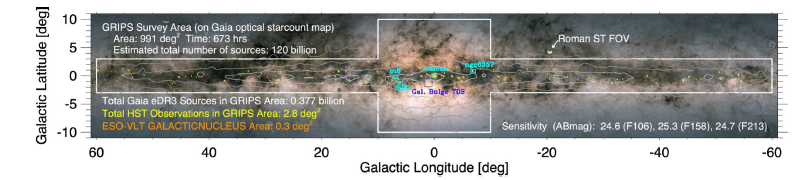}
\caption{Proposed observational plan for a Galactic Roman Infrared Plane Survey (GRIPS) overlaid on an optical Gaia starcount map which shows the high level of extinction in the Galactic Plane. The survey area is shown in white; the grey contours of COBE/DIRBE 4.9 $\mu$m emission (at 1,2,4,8 and 16 MJy/sr), which outline the stellar disk and bulge, fit within the urvey area. Of the 991 deg$^{2}$ covered, only 2.8 deg$^{2}$ / 0.3 deg$^{2}$ has high angular resolution coverage by HST *yellow)/ESO-VLT GALACTICNUCLEUS program (red). Approximate area of the 1.96 deg$^{2}$ Roman Galactic Bulge Time Domain Survey is shown in blue. For example star-forming fields -- Galactic Center, NGC 6357, M8 and M20 -- are noted in cyan. The combination of lowered extinction in the K band with high angular resolution will allow for the identification of 120 billion sources, a significant fraction of the Milky Way's stars. For comparison, this area contains only 0.38 billion sources in Gaia eDR3.}
\end{figure}

\begin{figure}
\centering
\includegraphics[width=1\linewidth]{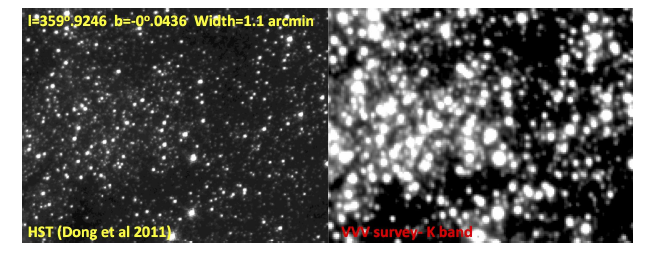}
\caption{HST 1.9 $\mu$m (F190) Galactic Center image from Dong et al. (2011, left) compared to VVV K$_{s}$ image (MInniti et al. 2010, right) of the same region. GRIPS will combine the ability to see through high dust columns with high angular resolution.}
\end{figure}

\begin{figure}
\centering
\includegraphics[width=1\linewidth]{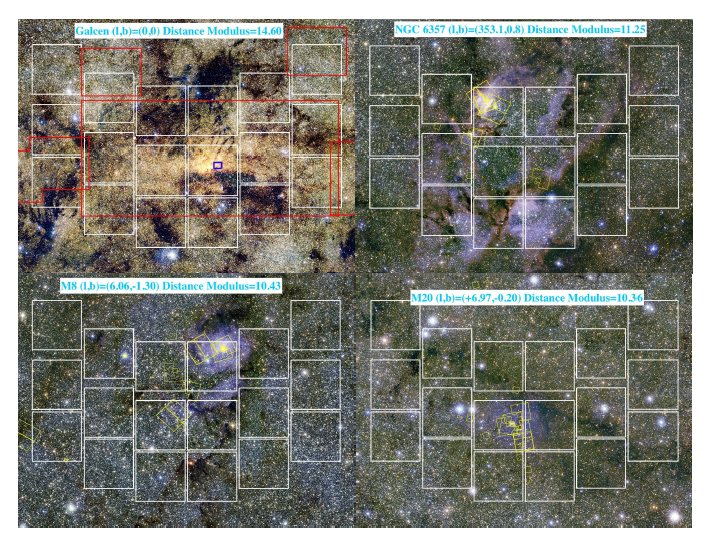}
\caption{VVV JHK$_{s}$ mosaics of four (out of thousands) Galactic star-forming fields in the GRIPS survey area with the Roman footprint (white), previous HST (yellow), and ESO-VLT (red) coverage superposed. A zoom-in of the violet region in the Galactic Center panel is shown in Fig.~2. With a K-band sensitivity limit of 22.85 (Vega) mags and assuming no extinction/source confusion, Roman/GRIPS can reach M$_{Ks}$ 8.25 (M5V) for sources at the distance of the Galactic Center and M$_{Ks}$ = 12.5 (mid L dwarfs) for sources in M8 and M20. A solar type star could be detected on the far side of the Galactic disk, even with 2-3 magnitudes of absorption.}
\end{figure}

\begin{figure}
\centering
\includegraphics[width=1\linewidth]{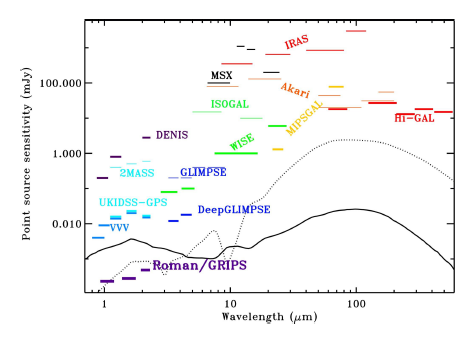}
\caption{Comparison of the point source sensitivity of the Roman/GRIPS survey with previous infrared all-sky and Galactic Plane surveys. The curves show model spectra of 1 L$_{\odot}$ T Tauri star at a distance of 14 kpc (solid curve) and a deeply embedded 1 L$_{\odot}$ protostar at a distance of 12 kpc (dotted curve.}
\end{figure}

\end{document}